\documentclass[amsmath,amssymb,showkeys,12pt,tightenlines,nobibnotes,nofootinbib]{revtex4}
\usepackage{amsfonts}
\usepackage{mathrsfs}
\usepackage{graphicx}
\usepackage{amsmath}
\usepackage{amssymb}
\usepackage{amsbsy}
\usepackage{marvosym}
\usepackage{url}
\makeatletter
\newcommand{\linea}{\noindent\rule{1.0\textwidth}{1pt}}%

\usepackage{color}
\makeatother

\begin{document}

\title{Learning by random walks in the weight space of the Ising perceptron}

\author{Haiping Huang$^{1}$}
\affiliation{$^1$Key Laboratory of Frontiers in Theoretical Physics,
Institute of Theoretical Physics, Chinese Academy of Sciences,
Beijing 100190, China\\
$^2$Kavli Institute for Theoretical Physics China, Institute of
Theoretical Physics, Chinese Academy of Sciences, Beijing 100190,
China}
\author{Haijun Zhou$^{1,2}$}
\affiliation{$^1$Key Laboratory of Frontiers in Theoretical Physics,
Institute of Theoretical Physics, Chinese Academy of Sciences,
Beijing 100190, China\\
$^2$Kavli Institute for Theoretical Physics China, Institute of
Theoretical Physics, Chinese Academy of Sciences, Beijing 100190,
China}
\date{\today}

\begin{abstract}
Several variants of a stochastic local search process for
constructing the synaptic weights of an Ising perceptron are
studied. In this process, binary patterns are sequentially presented
to the Ising perceptron and are then learned as the synaptic weight
configuration is modified through a chain of single- or
double-weight flips within the compatible weight configuration space
of the earlier learned patterns. This process is able to reach a
storage capacity of $\alpha \approx 0.63$ for pattern length $N =
101$ and $\alpha \approx 0.41$ for $N = 1001$. If in addition a
relearning process is exploited, the learning performance is further
improved to a storage capacity of $\alpha \approx 0.80$ for $N =
101$ and $\alpha \approx 0.42$ for $N=1001$. We found that, for a
given learning task, the solutions constructed by the random walk
learning process are separated by a typical Hamming distance, which
decreases with the constraint density $\alpha$ of the learning task;
at a fixed value of $\alpha$, the width of the Hamming distance
distributions decreases with $N$.
\end{abstract}

 \maketitle

\section{Introduction}
\begin{center}
\begin{figure}
         \includegraphics[bb=107 560 474 762,width=7.5cm]{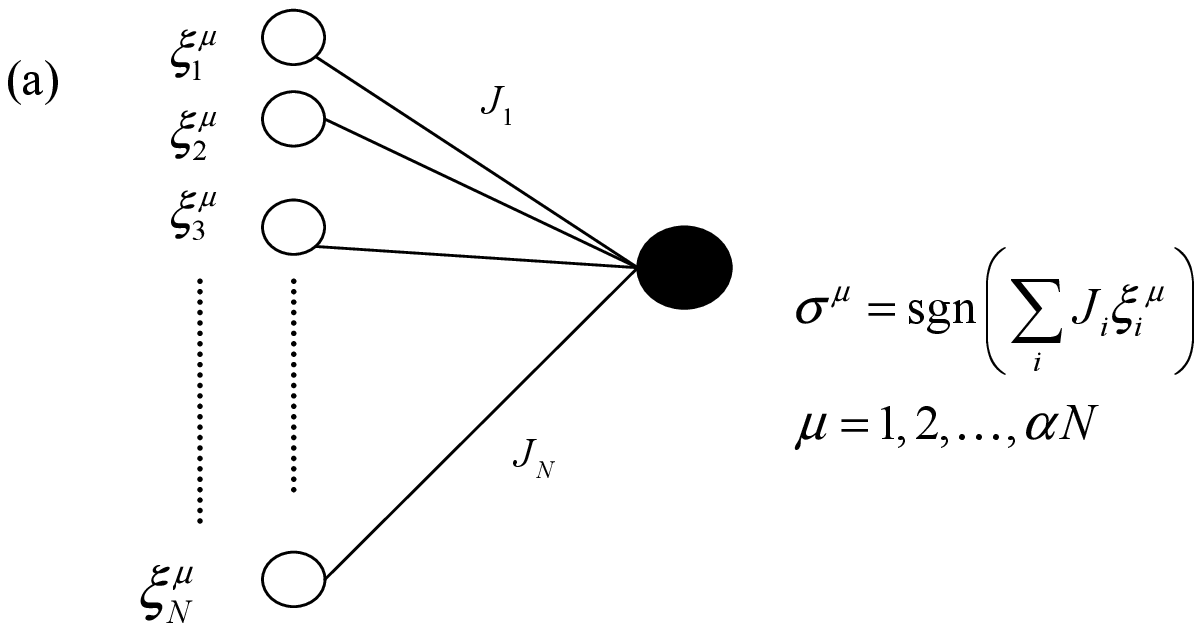}
     \hskip .5cm
     \includegraphics[bb=92 668 380 774,width=7.5cm]{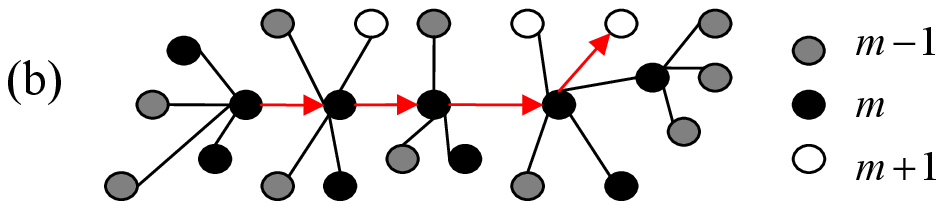}\vskip .2cm
  \caption{(Color online)
     The sketch of the Ising perceptron and the single-weight random walking process in the
     corresponding weight space. (a) $N$ input units (open circles)
     feed directly to a single output unit (solid circle). A binary input pattern
    $(\xi_1^\mu, \xi_2^\mu, \ldots, \xi_N^\mu)$ of length $N$
    is mapped through a sign function to a binary output $\sigma^{\mu}$, i.e.,
    $\sigma^\mu = {\rm sgn}\bigl(\sum_{i=1}^{N} J_{i}\xi_{i}^{\mu}\bigr)$. The set of $N$
      binary synaptic weights $\{J_{i}\}$ is regarded as a solution of the perceptron
     problem if the output $\sigma^\mu= \sigma_0^\mu$ for each of the $M= \alpha N$ input patterns $\mu \in [1, M]$, where
     $\sigma_0^\mu$ is a preset binary value.
      (b) A solution space random walking path (indicated by arrows).
      An open circle represents a configuration that satisfies the first $m+1$ input patterns, while
      a black circle and a gray circle represents, respectively, a configuration that satisfies
      the first $m$ and the first $m-1$ input patterns.
      An edge between two configurations means that
      these two configurations are related by a single-weight flip.
   }\label{perc}
 \end{figure}
 \end{center}

A single-layered feed-forward network of neurons, referred to as a
perceptron, is an elementary building block of complex neural
networks. It is also one of the basic structures for learning and
memory~\cite{Engel-2001}. In a perceptron, $N$ input neurons (units)
are connected to a single output unit by synapses of continuous or
discrete-valued synaptic weights. The learning task is to set the
weight values for these $N$ synapses such that an extensive number
$M=\alpha N$ of input patterns are correctly classified (see
Fig.~\ref{perc}a). The parameter $\alpha \equiv M/N$ is called the
constraint density. An assignment of these weights is referred to
as a solution if the perceptron correctly
classifies all the input patterns with this weight assignment. Compared with
perceptrons with real-valued synaptic weights, Ising perceptrons,
whose synaptic weights are binary, are much simpler for large-scale
electronic implementations and more robust against noise.
 An Ising perceptron is also relevant in real neural systems,
as the synaptic weight between two neurons actually takes bounded
values and has a limited number of synaptic
states\cite{Hopfield-1998,Wang-2005}. On the other hand, training a
real-valued perceptron is easy (e.g., the Minover algorithm
\cite{Krauth-1987} and the AdaTron algorithm \cite{Adatron-1989})
but training an Ising perceptron is known to be an NP-complete
problem \cite{Fusi-05}. Given $\alpha N$ input patterns, the
computation time needed to find a solution may grow exponentially
with the number of weights $N$ in the worst case. A complete
enumeration of all possible weight states is only feasible for small
systems up to
$N=25$~\cite{Krauth-1989b,Gutf-1990,Derrida-1991,Mona-1992}. In
recent years researches on efficient heuristic algorithms were
rather active
\cite{Fusi-05,Kohler-1990,Kohler-1990b,Bouten-1996,Kobe-1997,Zecchina-2006,Baldassi-2007,Baldassi-2009}.

If the number $M$ of input patterns is too large, a perceptron
will be unable to correctly classify all of them, no matter how the
synaptic weights are modified. This is a phase transition phenomenon of
the solution space of the perceptron.
In the case that the $M$ input binary patterns are sampled
uniformly and randomly from the set of all binary patterns,
the maximal value $\alpha_s$ of the constraint density $\alpha$,
the storage capacity at which a solution still exists,
has been calculated by statistical
physics methods. For the continuous perceptron subject to the spherical
constraint, Gardner and Derrida found that $\alpha_s=2$
\cite{Gardner-1988a,Gardner-1988b}.  At the thermodynamic limit of
$N\rightarrow \infty$, the continuous perceptron is impossible to
correctly classify more than $2 N$ random input patterns. When the
synaptic weight is restricted to binary values, $\alpha_s$
was predicted to be $0.83$ by Krauth and
M{\'{e}}zard using the first-step replica-symmetry broken spin-glass
theory \cite{Krauth-1989}. This prediction was confirmed by
numerical simulations of small size systems (plus an extrapolation
to large $N$) \cite{Gardner-1989,Krauth-1989b,Derrida-1991}.

The theoretically predicted storage capacity $\alpha_s$ represents the
upper limit of achievable constraint density $\alpha$ by any learning
strategies. As the constraint density $\alpha$ increases, it is expected
that the solution space of the Ising perceptron breaks into a huge
number of disjoint ergodic components \cite{Horner-1992a}. Solutions
from different components are significantly different. One can
define a connected component of the weight space as a cluster of
solutions in which any two solutions are connected by a path of
successive single-weight flips \cite{Arde-08,Kaba-09}. These
solution clusters are separated by weight configurations that only
correctly classify a subset of the input patterns. These partial
solutions act as dynamical traps for local search algorithms and make
the learning task  hard.
 An adaptive genetic algorithm was suggested by K\"ohler
in 1990, which could reach $\alpha \simeq 0.7$ for systems of
$N=255$ \cite{Kohler-1990}. Simulated annealing techniques were used
by Horner \cite{Horner-1992a} but critical slowing down of the
search process was observed, due to the very rugged energy landscape
of the problem. The simulated annealing was also used to study the
statistical structure of the energy landscape for the Ising
perceptron. The analysis of the distribution of distances between
global minima obtained by simulated annealing for small $\alpha$
indicated that the distance distribution becomes
a delta function in the thermodynamic
limit \cite{Fonta-1990}. Making use of the advantage that
efficient algorithms exist for the real-valued perceptron, an
alternative approach was to clip the trained real-valued weights of
the continuous perceptron into binary
values~\cite{Bouten-1996,Bouten-1998,Penney-1993,Penney-1993b,Malzahn-2000}.
Not all synaptic weights can be correctly specified by clipping,
however, and for those uncertain weights, complete enumeration was
then adopted. A message-passing algorithm was developed by Braunstein and Zecchina
for the Ising perceptron \cite{Zecchina-2006}, which
was able to reach $\alpha \simeq 0.7$ for $N\geq 1000$. The
efficiency of this belief-propagation algorithm was later
conjectured to be due to the existence of a sub-exponential number
of large solution clusters in the weight space \cite{Kaba-09}. An
on-line learning algorithm inspired from this belief-propagation
algorithm was also studied \cite{Baldassi-2007}, in which hidden
discrete internal states are added to the synaptic weights.

In real neural systems, the microscopic mechanism of perceptronal
learning is the Hebbian rule of synaptic modification
(spiking-time-dependent synaptic plasticity may be exploited, see,
e.g., Refs.~\cite{Gerstner-1999,DSouza-etal-2010}). The learning
processes in biological perceptronal systems are expected to be much
simpler than the various sophisticated learning processes of
artificial perceptrons. Two other important aspects of biological
perceptron systems are (i) the patterns to be classified are usually
read into the system in a sequential order, so they are being
learned one by one, and (ii) when a new pattern is being learned,
there are biological mechanisms which reactivate old learned
patterns; such recalling processes help to prevent old patterns from
being forgot as new patterns are learned (see, e.g., the
experimental investigation of Ref.~\cite{Kuhl-etal-2010}). Motivated
by these biological considerations, we investigate in this paper a
simple sequential learning mechanism, namely synaptic-weight space
random walking. In this random walking mechanism, the $\alpha N$
patterns are introduced into the system in a randomly permuted
sequential order, and random walk of single- or double-weight flips
is performed until each newly added pattern is correctly classified
(learned). The previously learned patterns are {\em not} allowed to
be misclassified in later stages of the learning process. We perform
extensive numerical simulations on several variants of this simple
sequential local learning rule and find that this mechanism has good
performance on systems of $N\sim 10^3$ neurons or less.

 The paper is organized as follows. The Ising perceptron learning is defined
in more detail in Sec.~\ref{sec_RCP}. Several strategies of learning
by random walks are presented in Sec.~\ref{sec_LRW}. In
Sec.~\ref{sec_result}, experimental study of learning algorithms is
carried out. The overlap distribution of solutions as well as
performances of different local search algorithms is reported.
Summary and discussion are given in Sec.~\ref{sec_Sum}.

Sequential random walk search algorithms were recently investigated
in various combinatorial satisfaction problems (see, e.g.,
Refs.~\cite{Arde-07,Zhou-2009a,Zhou-2009}). The present
work adds evidence that the solution space random walking mechanism,
although very simple and easy to implement, is able to solve many
nontrivial problem instances of a given complex learning or
constraint satisfaction problem.

\section{The random classification problem}
\label{sec_RCP}

For the Ising perceptron depicted schematically in Fig.~\ref{perc}a,
$N$ input units are connected to a single output unit by $N$
synapses of weight $J_i =  \pm 1$ $(i = 1, 2, \ldots, N)$. The
perceptron tries to learn $M = \alpha N$ associations $\{\xi^\mu ,
\sigma_0^\mu \}$ $(\mu= 1, 2, \ldots, M)$, where $\xi^\mu \equiv
(\xi_1^\mu, \xi_2^\mu, \ldots, \xi_N^\mu)$ is an input pattern with
$\xi_i^\mu = \pm 1$, and $\sigma_0^\mu = \pm 1$ is the desired
classification of the input pattern $\mu$. Given the input pattern
$\xi^\mu$, the actual output $\sigma^\mu$ of the perceptron is
\begin{equation}
    \sigma^\mu = {\rm sgn}\Bigl( \sum\limits_{i=1}^N J_i \xi_i^\mu
    \Bigr) \ .
\label{output}
\end{equation}
The perceptron can modify its synaptic weight configuration $\{J_i\}
\equiv ( J_1, J_2, \ldots, J_N)$ to achieve complete classification,
i.e., $\sigma^\mu = \sigma_0^\mu$ for each of the $M$ input pattern.
The solution space of the Ising perceptron is composed of all the
weight configurations $\{J_i\}$ that satisfy  $\sigma_0^\mu \sum_i
J_i \xi_i^\mu > 0$ for $\mu = 1, 2, \ldots, M$.

For the random Ising perceptron problem studied in this paper, each
of the $M$ input binary patterns $\xi^\mu$ is sampled uniformly and
randomly from the set of all $2^N$ binary patterns of length $N$,
and the classification $\sigma_0^\mu$ is equal to $\pm 1$ with equal
probability. For $N$ sufficiently large, the solution space of such a
model system is non-empty as long as $\alpha  < 0.83$
\cite{Krauth-1989}. To
construct such a solution configuration $\{J_i\}$, however, is quite
a non-trivial task.

A more stringent learning problem is to find a weight configuration
$\{J_i\}$ such that, for each input pattern $\xi^\mu$,
\begin{equation}
    \sigma_0^\mu \frac{ \sum_i J_i \xi_i^\mu }{\sqrt{N}} \geq \kappa \ ,
    \label{eq:kappa}
\end{equation}
where $\kappa > 0$ is a preset parameter \cite{Krauth-1989}. The
most efficient way of solving this constraint satisfaction problem
appears to be the message-passing algorithm of
Refs.~\cite{Zecchina-2006,Baldassi-2007}.

One can perform a gauge transform of $\xi_i^\mu \rightarrow
\xi_i^\mu \sigma_0^\mu$ to each input pattern. Under this gauge
transform, each desired output is transformed to $\sigma_0^\mu = 1$.
Without loss of generality, in the remaining part of this paper we
will assume $\sigma_0^\mu = 1$ for any input pattern $\mu$. Consider
the case of $N$ being odd, we define the stability field of a
pattern $\mu$ as
\begin{equation}
 h^\mu = \sum\limits_{i=1}^{N} J_i \xi_i^\mu \ .
\end{equation}
To ensure the local stability of input pattern $\mu$ under changes
of weight configuration $\{J_i\}$, in analogy to
Eq.~(\ref{eq:kappa}),  we introduce a stability parameter $\Delta
\geq 1$ and require that $h^\mu \geq \Delta $ for each $\mu$. Input
patterns with $h^\mu \geq 3$ are stable against a single-weight
flip. For the single-weight flipping processes of the next section,
the input patterns with $h^\mu = 1$ are referred to as barely
learned patterns, as these patterns may become misclassified after
the weight configuration makes a single flip. Similarly, for the
double-weight flipping process of the next section, the input
patterns with $h^\mu = 1$ or $h^\mu=3$ are referred to as barely
learned patterns.

\section{Learning by random walks}
\label{sec_LRW}

Random walk processes were used in a series of works
 \cite{Arde-07,Zhou-2009a,Cocco03,Semer-03,Bart-03,MonaRev08} to
find solutions
 for constraint satisfaction problems. They were also used as tools
 to study the
solution space structure of these constraint satisfaction problems
\cite{Arde-07,Zhou-2009a,Krza-07}. Various local search strategies
have been developed to improve the performance of random walk
 stochastic searching~\cite{Seitz-05,Arde-06,Alava-2008}.

 The random walk learning strategies of this work follow the {\tt SEQSAT}
 algorithm
 of Ref.~\cite{Zhou-2009a}.
An initial weight configuration $(J_1^{(0)}, J_2^{(0)}, \ldots,
J_N^{(0)})$ is randomly generated at time $t=0$. The first pattern
$\xi^1$ is applied to the Ising perceptron. If this
 pattern is correctly classified under the initial weight configuration
(i.e., $h^1 > 0$),
 then the second pattern $\xi^2$ is applied; otherwise
the weight configuration is adjusted by a sequence of elementary
local changes until $\xi^1$ is correctly classified. The algorithm
then proceeds with the second pattern $\xi^2$,
 the third pattern $\xi^3$, etc., in a sequential order. An elementary
local change of weight configuration is achieved either by a
single-weight flip (SWF) or by a double-weight flip (DWF).

Suppose at time $t$ the weight configuration is $\{J^{(t)}\}\equiv
(J_1^{(t)}, J_2^{(t)}, \ldots, J_N^{(t)} )$, and suppose this
configuration correctly classifies the first $m$ input patterns
$\xi^\mu$ ($\mu = 1, \ldots, m)$ but not the $(m+1)$-th pattern
$\xi^{m+1}$. The configuration $\{J_i\}$ will keep wandering in the
solution space of the first $m$ patterns until a configuration that
correctly classifies $\xi^{m+1}$ is reached (see Fig.~\ref{perc}b).
In the SWF protocol, a set $A(t)$ of allowed single-weight flips is
constructed based on the current configuration $\{J^{(t)}\}$ and the
$m$ learned patterns. $A(t)$ includes all integer positions $j \in
[1, N]$ with the property that the single-weight flip of $J_j^{(t)}
\rightarrow -J_j^{(t)}$ does not render any barely learned patterns
$\mu \in [1, m]$ (whose $h^\mu = 1$) being misclassified. At time
$t^\prime= t+ 1/N$
 an integer position
$j$ is chosen uniformly and randomly from set $A(t)$ and the weight
configuration is changed to $\{J^{(t^\prime)}\}$ such that
$J_i^{(t^\prime)}= J_i^{(t)}$ if $i \neq j$ and
$J_j^{(t^\prime)}=-J_j^{(t)}$. It is obvious that the new
configuration $\{J^{(t^\prime)}\}$ also satisfies all the first $m$
patterns.

The DWF protocol is very similar to the SWF protocol, with the only
difference that the allowed set $A(t)$ at time $t$ contains ordered
pairs of integer positions $(i, j)$ with $i < j$. This set of
ordered pairs can also be easily constructed. If, with respect to
configuration $\{J^{(t)}\}$, there are no barely learned patterns
(whose stability field $h^\mu = 1$ or $3$) among the first $m$ learned
patterns, then $A(t)$ contains all the $N(N-1)/2$ ordered pairs of
integers $(i, j)$ with $1 \leq i < j \leq N$. Otherwise, randomly
choose a barely learned pattern, say $m_1 \in [1, m]$, and for each
integer $i \in [1, N]$ with the property that $J_i^{(t)}
\xi_i^{m_1}< 0$, do the following: (1) if $J_i^{(t)} \xi_i^{\mu} <
0$ for all the other barely learned patterns, then add all the
ordered pairs $(i, j)$ with $j\in [i+1, N]$ into the set $A(t)$; (2)
otherwise, add all the ordered pairs $(i, j)$ into the set $A(t)$,
with the property that the integer $j \in [i+1, N]$ satisfies
$J_j^{(t)} \xi_j^{\mu} <
0$ for all those barely learned patterns $\mu\in [1, m]$ with
$J_i^{(t)} \xi_i^{\mu}>0$.

The waiting time $\Delta t_{m+1}$ of satisfying the $(m+1)$-th
pattern is defined as the total elapsed time from first satisfying
the $m$-th pattern to first satisfying the $(m+1)$-th pattern. And
the total time $T_{m+1}$ of satisfying the first $(m+1)$ patterns is
simply $T_{m+1}= \sum_{\mu=1}^{m+1}
 \Delta t_{\mu}$. One time unit corresponds to $N$ elementary local changes of
the weight configuration. The random walk searching process stops if
all the $M$ input patterns have been correctly classified, or if the
last visited weight configuration becomes an isolated point (i.e.,
the set $A(t)$ becomes empty after a new pattern is included into the
set of learned patterns), or if the
last waiting time $\Delta t_{m+1}$ exceeds a preset maximal time
value $\Delta t^{max}$, which is equal to $\Delta t^{max}= 1000$
  in the present work.

The SWF and DWF random walks processes as mentioned above are very
simple to implement and they do not overcome
any barriers in the energy landscape of the perceptron learning problem.
However, as we demonstrate in the next section, their performances are
quite remarkable for problem instances with pattern length
$N\leq 10^3$.

The SWF process, as a local search algorithm, will get stuck in one
of the enormous metastable states when all the weights become frozen
(here we identify a synaptic weight as being frozen if flipping its
value causes at least one of the learned patterns to be
misclassified), at a constraint density value much smaller than the
theoretical threshold value of $0.83$. The DWF process will also get
jammed if the weight configuration becomes frozen with respect to
any double-weight flips. To further improve the achievable storage
capacity for the SWF and DWF learning processes, a simple relearning
strategy is added to the random walk searching. The basic idea of
the relearning strategy is: if some learned patterns are hindering
the learning of new patterns very much, we first ignore them and
proceed to learn a number of new patterns; after that, we learn the
ignored patterns again and hope they can all be correctly
classified.

In the present work, we implement the relearning strategy in the following
way. Suppose that as the $m$-th input pattern is presented to the Ising
perceptron, the SWF or the DWF process is unable to learn it in a waiting
time $\Delta t_{m} < \Delta t^{max}$. We then remove all the
$k$ barely learned patterns $\mu\in [1, m-1]$ with $h^\mu =1$ from the
list of learned patterns, and proceed to learn
 the patterns $\mu \in [m, m+k-1]$ in a sequential manner
(stage 1). If the SWF process
or the DWF process succeeds in learning these $k$ patterns,
 we then return to learn the $k$ previously
removed patterns again in a sequential manner (stage 2). If this
relearning succeeds, we proceed with the patterns with index $\mu
\geq m+k$. If this attempt fails either at stage $1$ or at stage
$2$, we stop the whole random walk learning process or start with
another trial by removing all the learned patterns. In practice, we
find that the relearning process has a high probability to succeed
in both stage $1$ and stage $2$ if $\alpha$ is not too large and
pattern length is of order $10^{3}$ or less.

\section{Results}
\label{sec_result}

\begin{figure}
    \includegraphics[width=0.7\textwidth]{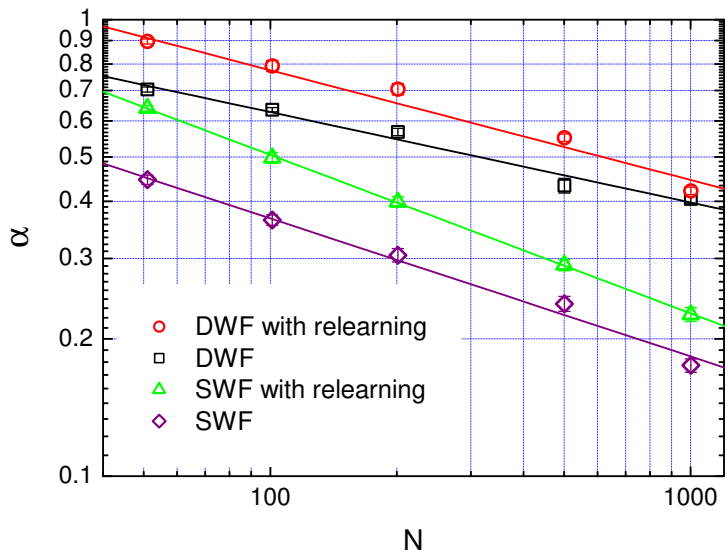}
  \caption{
  (Color online) Comparison of the performances of several random walk search strategies.
  The achieved storage capacity $\alpha$
 as averaged over many independent runs
 ($100$ for the smallest $N$ and $10$ for the largest $N$) are
 shown as a function of the pattern length $N$.
  The solid lines are power-law fittings of the form $\alpha \propto N^{-\gamma}$,
  with $\gamma = 0.302, 0.347, 0.198, 0.241$ for SWF, SWF with relearning, DWF and DWF with relearning, respectively.
  }\label{CapN}
\end{figure}

Figure~\ref{CapN} demonstrates the simulation results for several
random walk learning strategies. For each learning strategy,
$\mathcal{N}$  set of random input patterns $(\xi^1, \xi^2, \ldots,
\xi^M)$ are generated. Each input pattern $\xi^\mu$ has length $N$.
The random walk
learning strategy is then applied to each set of patterns until it
stops, at which point we record the number of correctly classified
patterns $m$ and calculate the achieved
storage capacity $\alpha = m /N$. The
mean values of $\alpha$  are reported in Fig.~\ref{CapN}. It appears
that the storage capacity of all the four learning strategies
decreases with $N$ roughly as  a power law $\alpha \propto
N^{-\gamma}$. At each value of $N$, the SWF strategy has the worst
performance, while the DWF strategy with relearning has the best
performance.

The SWF strategy is able to reach a storage capacity of $\alpha
\approx 0.36$ for systems of $N=101$ and $\alpha \approx 0.17$ for
systems of $N=1001$. These values are much less than the theoretical
storage capacity of $\alpha \approx 0.83$. However, the DWF
strategy performs much better, with a capacity of $\alpha \approx
0.63$ for $N=101$ and $\alpha \approx 0.41$ for $N=1001$. In real
neural systems, perceptronal learning of elementary patterns
probably does not involve too many neuronal cells and a value of $N
\sim 10^2$ might be common. For perceptronal systems with $N \sim
10^2-10^3$, the SWF and DWF strategies can be regarded as efficient.

If relearning is introduced into the random walk learning
strategies, the performance can be further improved. For the DWF
strategy with relearning, we find that the storage capacity is
$\alpha \approx 0.80$ for $N=101$ and $\alpha \approx 0.42$ for
$N=1001$. Relearning is indeed a biologically relevant strategy in
perceptronal learning of real neural systems
\cite{Kuhl-etal-2010,Abbott-2007}. As a comparison, for problem
instances of pattern length $N=1001$, the belief-propagation
inspired learning strategy of Baldassi and coauthors
\cite{Baldassi-2007} achieves $\alpha \approx 0.47$ when the
number $K$ of internal states of their algorithm is set to $K=40$.
This storage capacity $\alpha$ decreases to $\alpha \approx
0.36$ at $K=20$ and to $\alpha \approx 0.10$ at $K=10$.

For the same set of input patterns $(\xi^1, \xi^2, \ldots, \xi^m)$,
different runs of the SWF strategy or the DWF strategy lead to
different solution configurations. The similarity between solutions
can be measured by an overlap value $q$ as defined by
\begin{equation}\label{overlap}
        q=\frac{1}{N}\sum\limits_{i=1}^{N}J_{i}J_{i}^{\prime} \ ,
\end{equation}
where $(J_1, \ldots, J_N)$ and $(J_1^\prime, \ldots, J_N^\prime)$
are two solutions. The reduced Hamming distance $d_H$ between two
solutions is related to the overlap $q$ by $d_H = (1-q)/2$. The
typical value of the overlap value at constraint density $\alpha
\sim 0.83$ is predicted to be $q \approx 0.56$ according to the
replica-symmetric calculation \cite{Krauth-1989}, suggesting that
solutions are still far away from each other (with a reduced Hamming
distance $d_H \approx 0.22$) as $\alpha$ approaches the theoretical
storage capacity $\alpha_s$.

\begin{figure}
\centering
    \includegraphics[bb=9 16 286 219,width=0.8\textwidth]{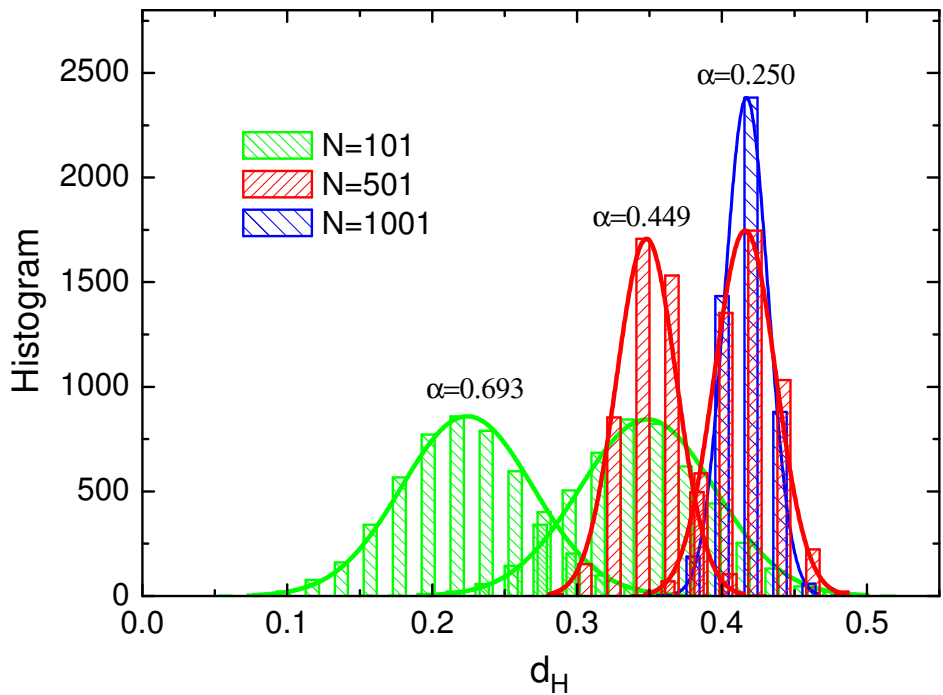}
  \caption{
    (Color online) Histograms of reduced Hamming distances between
    solutions found by DWF on a single problem instance of $M$ input
patterns of length $N$. $100$
    solutions are constructed for each of the five instances with
    $(M, N) = (45, 101)$, $(70, 101)$, $(125, 501)$, $(225, 501)$,
$(250, 1001)$,
    respectively. The solid lines are Gaussian fitting results to the
histograms.
}\label{Histo}
\end{figure}

Figure~\ref{Histo} shows the histogram $P(d_H)$ of reduced Hamming
distances $d_H$ between different solutions found by the DWF
strategy for a single problem instance with constraint density
$\alpha$ and pattern length $N$.
Different pattern lengths of $N= 101, 501, 1001$ are used,
and $100$ different solutions are constructed by repeated running of
the DWF process. Other problem instances show similar properties. We
notice from Fig.~\ref{Histo} that, at the same value of $\alpha$,
the histograms $P(d_H)$ for different $N$ are peaked at almost the
same $d_H$ value, but the width of $P(d_H)$ decreases as $N$ is
enlarged. Such a behavior was observed earlier in Ref.~\cite{Fonta-1990}
on a slightly modified Ising perceptron problem. The solutions obtained by the DWF strategy
therefore have a typical level of similarity. Figure~\ref{Histo}
also demonstrates that, as the constraint density increases, the
histograms $P(d_H)$ shift to smaller $d_H$ values, suggesting that
the level of similarity between the DWF-constructed solutions
increases with $\alpha$. At $\alpha=0.693$ the typical reduced
Hamming distance is $d_H \approx 0.224$, compatible with the
mean-field predictions \cite{Krauth-1989}. Similar results are
obtained for solutions found by the SWF strategy. In all our
simulations, we do not observe double or multiple peaks for the
histogram $P(d_H)$. The results of these and our other numerical
simulations (not shown) are consistent with proposal that, for a
given problem instance, the solutions obtained by the random walking
strategies are members of the same (large) solution cluster of the
solution space~\cite{Fonta-1990,Biro-00,Kaba-09}. Unlike the random
$K$-satisfiability problem, the random $Q$-coloring problem, or
some locked constraint satisfaction problems
\cite{Krzakala-etal-PNAS-2007,Zhou-2009b,Lenka-08}, the solution
space organization of the Ising perceptron problem is still not very
clear. Kabashima and co-authors \cite{Kaba-09} suggested that for
$\alpha<0.83$ the solution space of the Ising perceptron
problem is equally dominated by exponentially many clusters of
vanishing entropy and a sub-exponential number of large clusters.
Our simulation results are compatible with this proposal, but
more work needs to be done to clarify the solution space structure of
the random Ising perceptron problem.

\begin{figure}
\centering
    \includegraphics[width=0.7\textwidth]{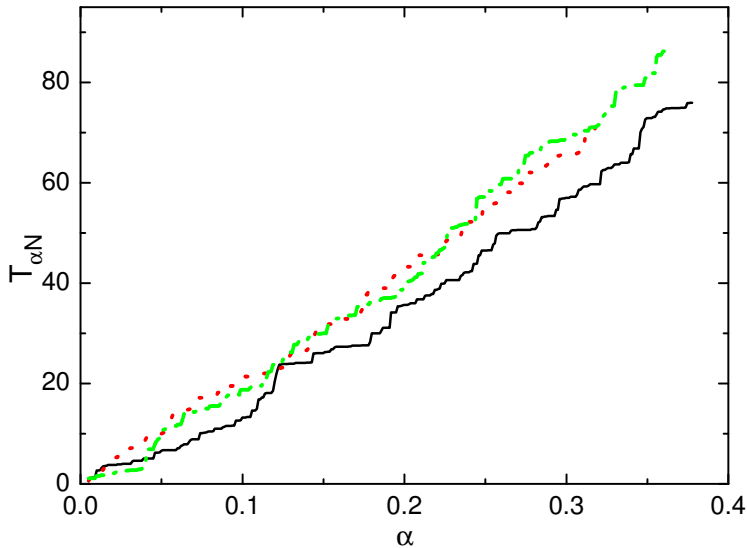}
  \caption{
    (Color online) The learning time $T_{\alpha N}$ as a function of $\alpha$
 for for three problem instances of $N=1001$.
  }\label{FracFroz}
\end{figure}

 The total time $T_{\alpha N}$ used by the DWF strategy to correctly classify
the first $\alpha N$ patterns for a problem instance with
$N=1001$ is shown in
Fig.~\ref{FracFroz} as a function of $\alpha$. The learning time
grows almost linearly with $\alpha$ for $\alpha < 0.4$.
As the constraint density $\alpha$ becomes large, different solution
communities are expected to form in the solution space
\cite{Zhou-2009b}. Then as $\alpha$ further increases to certain larger
value, the time needed for the random walk process to escape from a
solution community may exceed the preset
maximal waiting time of $\Delta t^{max}=1000$ and the DWF process will
then stop. The achieved storage capacity $\alpha$ can be increased to
some extent if we make $\Delta t^{max}$ larger, but the search process
will become more and more viscous as the solution space of the problem
becomes more and more heterogeneous and complex \cite{Zhou-2009a}.
We do not attempt to calculate the jamming point of the random walk
searching processes.

\section{Discussion}
\label{sec_Sum}

We proposed several stochastic learning strategies for the Ising
perceptron problem based on the idea of solution space random
walking \cite{Zhou-2009a}. Our simulation results in Fig.~\ref{CapN}
demonstrated that, the DWF strategy is able
to correctly classify $\geq 0.4 N$ random input patterns of length
$N$ for $N \leq 1001$. If a simple relearning strategy is added to
the DWF strategy, the learning performance is further improved. The
learning time of the DWF strategy grows roughly linearly with the
number of input patterns.  This work suggested that learning by
local and random changes of synaptic weights is efficient for
perceptronal systems with $N\approx 10^2-10^3$ neurons.
 These local sequential learning strategies may be exploited in some biological perceptronal
systems. In real neuronal systems, the number $N$ of involved
neurons in an elementary pattern classification task may be of the
order of $N \sim 10^1-10^3$.

The solutions obtained by the DWF strategy for a given perceptronal
learning task are separated by a typical Hamming distance, which
reduces as the number of input patterns increases
(Fig.~\ref{Histo}). However, solutions are still far away from each
other even near to the critical capacity. We suspected that for the
problem instances studied in this paper, either the solution space
of the problems is ergodic as a whole, or the solutions reached by
the DWF strategies all belong to the same solution cluster of the
solution space. In our random walking setting, once all weights are
frozen, particularly for SWF, the current pattern with negative
stability field will be no longer learned since the current weight
configuration is isolated in the weight space (this weight
configuration is denoted as the completely frozen solution);
fortunately, DWF is able to go on even if all weights are frozen,
since flipping certain pairs of weights is still permitted from the
configuration where each single weight is not allowed to be flipped.
If these flippable pairs of weights do not exist, DWF will get trapped,
and the configuration is isolated once again. Actually, as the
constraint density $\alpha$ increases, many such isolated solutions
will show up, and SWF or DWF working by single- or double-weight
flips, is not capable of crossing energy barriers separating the
isolated solutions from those connected ones, which can be bypassed
to some extent using the relearning strategy which helps to escape
from these small clusters and makes SWF or DWF keep on exploring the
large cluster composed of exponentially many solutions.  For small
$\alpha$, replica symmetric ansatz is believed to give a good
description of the solution space of Ising
percetpron~\cite{Fonta-1990}. Up to $\alpha_{s}$, point-like
clusters will form and searching for the compatible weights becomes
more difficult\cite{Lenka-08}. It is desirable to have a theoretical
understanding on the structural evolution of the solution space of
the random Ising perceptron problem. How the dynamics of stochastic
local search algorithms is influenced by the solution space
structure of the random Ising perceptron is an important open issue.

Another interesting problem is the generalization problem
 where the
inputs-output associations are no longer uncorrelated but the
desired outputs are given by a teacher perceptron
 \cite{Baldassi-2009,Gyo-1990,Somp-1990,Horner-1992b}. The
student perceptron tries to learn the rule provided by the teacher.
After an enough amount of examples are presented to the student
perceptron, the student's weights should match those of the
 teacher, then the
network undergoes a first-order transition from poor to perfect
generalization \cite{Gyo-1990,Somp-1990}.
It is worthwhile to extend the current random walk strategies
to  analyze the generalization problem in Ising
perceptrons.

\section*{Acknowledgments}

This work was partially supported by the National Science Foundation
of China (Grant numbers 10774150 and 10834014) and the China
973-Program (Grant number 2007CB935903).



\end{document}